\documentclass[prl,showpacs,floatfix,twocolumn,byrevtex]{revtex4}
\usepackage{dcolumn}
\usepackage{graphicx}
\begin{document}

\bibliographystyle{apsrev}

\preprint{Draft version 3, not for distribution}
%
%
\title{Charge order, metallic behavior and superconductivity in
 \boldmath La$_{2-x}$Ba$_x$CuO$_4$ with $x=1/8$ \unboldmath}

\author{C. C. Homes}
\email{homes@bnl.gov}
\author{S. V. Dordevic}
\altaffiliation{Department of Physics, The University of Akron, Akron, OH
44325}
\author{G. D. Gu}
\author{Q. Li}
\author{T. Valla}
\author{J. M. Tranquada}
\affiliation{Condensed Matter Physics \& Materials Science Department,
Brookhaven National Laboratory, Upton, New York 11973}%
\date{\today}

%
%
\begin{abstract}
The {\it ab}-plane optical properties of a cleaved single crystal of
La$_{2-x}$Ba$_x$CuO$_4$ for $x=1/8$ ($T_c \simeq 2.4$~K) have been
measured over a wide frequency and temperature range. The low-frequency
conductivity is Drude-like and shows a metallic response with decreasing
temperature.  However, below $\simeq 60$~K, corresponding to the onset of
charge-stripe order, there is a rapid loss of spectral weight below about
40~meV.  The gapping of single-particle excitations looks
surprisingly similar to that observed in superconducting
La$_{2-x}$Sr$_{x}$CuO$_4$, including the presence of a residual Drude
peak with reduced weight; the main difference is that the lost spectral
weight moves to high, rather than zero, frequency, reflecting the absence
of a bulk superconducting condensate.

\end{abstract}
%
%
\pacs{74.25.Gz, 74.72.-h, 78.30.-j}%
\maketitle

%
%
%

It has been proposed that charge inhomogeneity, especially in the form of
stripes, is a phenomenon intrinsic to doped antiferromagnets such as cuprate
superconductors \cite{kivelson03,zaan01,mach89,cast95,vojt99}. Static spin
and charge stripe order has been observed by diffraction in a couple of
cuprate compounds \cite{ichikawa00,fujita04,abba05}; however, it appears to
compete with superconductivity \cite{ichikawa00}.  The electronic structure
of the stripe-ordered state remains to be clarified experimentally.  This is
an important issue, as it bears on the question of whether dynamic stripes
might be compatible with superconductivity, and possibly even necessary for
the high transition temperatures \cite{arrigoni04}.  Kivelson, Fradkin, and
Emery \cite{kivelson98} argued that straight, ordered stripes should develop
a charge-density-wave (CDW) order along the stripes, which would compete with
pairing correlations.  Castellani {\it et al.} \cite{castellani00} have
suggested that such commensurate stripes should be insulating.  Previous
investigations of stripe correlations using infrared reflectivity on
oriented, cut-and-polished crystals of La$_{1.88-y}$Nd$_y$Sr$_{0.12}$CuO$_4$
(LNSCO) \cite{tajima99,dumm02} and La$_{2-x}$Sr$_x$CuO$_4$ (LSCO)
\cite{dumm02,lucarelli03} have resulted in a certain amount of controversy
\cite{tajima03,lucarelli03b}.

In this Letter we report on the {\it ab}-plane optical properties of a {\it
cleaved} single crystal of La$_{2-x}$Ba$_x$CuO$_4$ (LBCO) for $x=1/8$. The
same crystal was used in a recent soft-X-ray resonant-diffraction study that
demonstrated charge order below 60~K \cite{abba05}. Working with a cleaved
surface greatly reduces the possibility of inadvertent surface misorientation
that can result in spurious effects \cite{tajima03}. The optical conductivity
initially shows a normal (for the cuprates) metallic response with
decreasing temperature. Below $\simeq 60$~K, the conductivity at very low
frequency appears to either remain roughly constant or increase slightly
as the temperature is reduced; however, there is a general loss of
spectral weight (the area under the conductivity curve) below $\simeq
300$~cm$^{-1}$, a consequence of the substantial reduction in the density
of carriers.  The missing spectral weight is redistributed to higher
frequencies, and is fully recovered only above about 2000~cm$^{-1}$.
Coinciding with the depression of the carrier density, the transition to
superconductivity is depressed to 2.4~K.  The gapping of the
single-particle excitations looks remarkably similar to that recently
reported for superconducting La$_{1.85}$Sr$_{0.15}$CuO$_4$
\cite{tajima05}, which is known to have a $d$-wave gap \cite{yosh03}.  The
presence of a residual Drude peak indicates that charge order is
compatible with a ``nodal metal'' state \cite{lee05}.


%
%
Single crystals of La$_{2-x}$Ba$_x$CuO$_4$ with $x=1/8$ were grown by the
floating zone method.  The sample used in this study had a strongly suppressed
bulk $T_c \simeq 2.4$~K as determined by magnetic susceptibility. The sample
was cleaved in air, yielding a mirror-like {\it ab}-plane face.  The {\it
ab}-plane reflectance has been measured at a near-normal angle of incidence
over a wide temperature and spectral range using an {\it in-situ} evaporation
technique \cite{homes93}. The optical properties are calculated from a
Kramers-Kronig analysis of the reflectance, where extrapolations are supplied
for $\omega \rightarrow 0, \infty$.  At low frequency, a metallic Hagen-Rubens
response is assumed $(R \propto 1-\omega^{1/2})$.  Above the highest-measured
frequency in this experiment the reflectance of La$_{1.85}$Sr$_{0.15}$CuO$_4$
has been employed to about 40~eV \cite{uchida91}; above that frequency a
free-electron approximation has been assumed $(R\propto 1/\omega^4)$.

\begin{figure}[t]%
%
%
\vspace*{-0.5cm}%
\centerline{\includegraphics[width=3.7in]{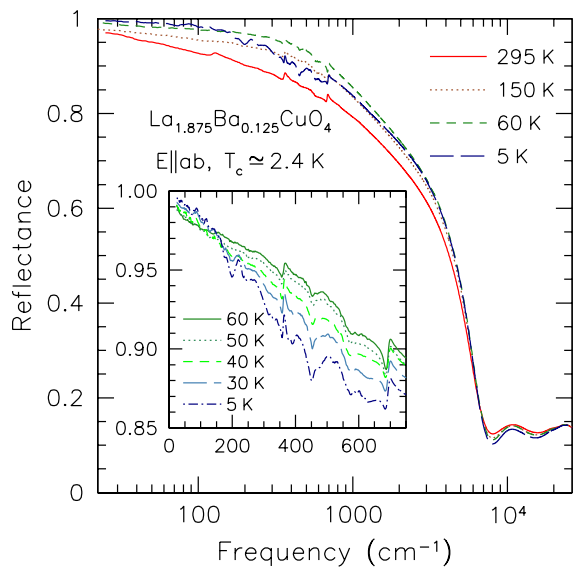}}%
\vspace*{-1.4cm}%
\caption{(Color online) The reflectance at a near-normal angle of
incidence for several temperatures over a wide spectral range for a
cleaved surface of La$_{1.875}$Ba$_{0.125}$CuO$_4$ ($T_c \simeq 2.4$~K)
for light polarized in the {\it a-b} plane. The infrared reflectance
increases with decreasing temperature until about 60~K, below which it
is suppressed in the $200 - 2000$~cm$^{-1}$ region.
Inset: The detailed temperature dependence of the far-infrared
reflectance at and  below 60~K. (The resolution is 2~cm$^{-1}$.)}
\vspace*{-0.3cm}%
\label{fig:reflec}
\end{figure}

%
%
The reflectance of La$_{1.875}$Ba$_{0.125}$CuO$_4$ is shown from $\approx 20$
to 25,000~cm$^{-1}$ in Fig.~\ref{fig:reflec} for a variety of temperatures. The
reflectance is typical of many cuprates, with a plasma edge at $\approx 1$~eV
and a low-frequency reflectance that increases with decreasing temperature,
indicative of a metallic response.  However, below $\approx 60$~K the
low-frequency reflectance in the $200 - 2000$~cm$^{-1}$ region stops increasing
and begins to decrease, as shown in  more detail in the inset of
Fig.~\ref{fig:reflec}. Note that below $\approx 180$~cm$^{-1}$ the reflectance
continues to increase below 60~K.  This suppression of the far-infrared
reflectance is unusual and has not been observed in other studies of LSCO or
LNSCO \cite{uchida91,gao93,startseva99,dumm02,lucarelli03, tajima05,padilla05}
for $T>T_c$.
%
%
%
%

\begin{figure}[t]%
%
%
\vspace*{-0.8cm}%
\centerline{\includegraphics[width=3.6in]{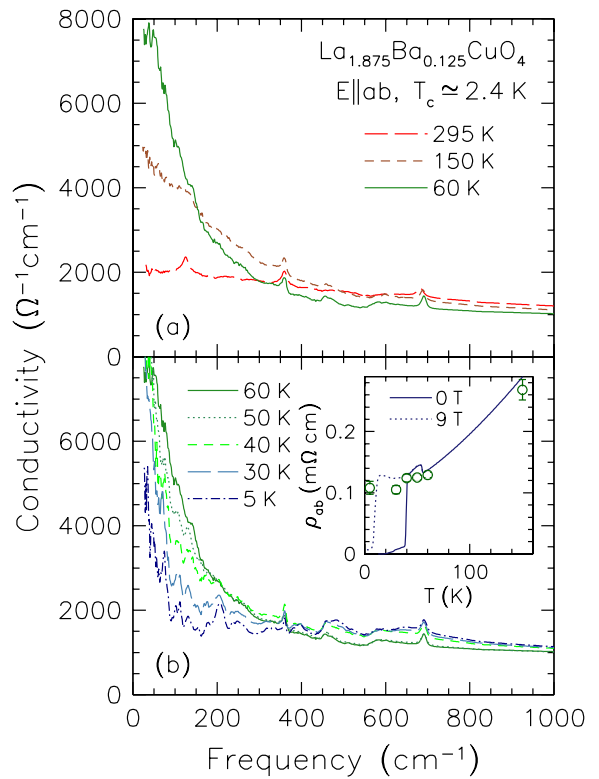}}%
\vspace*{-1.2cm}%
\caption{(Color online) The {\it ab}-plane optical conductivity of
La$_{1.875}$Ba$_{0.125}$CuO$_4$ ($T_c \simeq 2.4$~K). (a) The conductivity in
the infrared region between 295 and 60~K, showing a steady narrowing of the
Drude-like component.  The sharp features in the conductivity are the
normally-allowed infrared active vibrations; a feature at $\simeq
125$~cm$^{-1}$ at 295~K is not visible at 60~K.
(b) The conductivity for several temperatures below 60~K.  A new vibrational
feature appears at $\simeq 205$~cm$^{-1}$ below about 40~K.
Inset: The {\it ab}-plane resistivity of La$_{2-x}$Ba$_x$CuO$_4$ for $x
\simeq 1/8$ (solid curve) and in an 9~T magnetic field (dotted curve).  The
open circles are the estimated values for the dc resistivity calculated from
the Drude parameters in Table~\ref{tab:drude}. }
\vspace*{-0.3cm}%
\label{fig:sigma}
\end{figure}

The optical conductivity is shown in Fig.~\ref{fig:sigma}, with the
conductivity between 295 and 60~K in panel (a) and the behavior below 60~K in
(b).  The conductivity can be described as a combination of a coherent
temperature-dependent Drude component, and an incoherent
temperature-independent component.  The ``two-component'' expression for the
real part of the optical conductivity is
\begin{equation}
  \sigma_1(\omega) = {1\over{60}}\, { {\omega_{p,D}^2 \Gamma_D} \over
   {\omega^2+\Gamma_D^2} }+\sigma_{MIR},
\end{equation}
where $\omega_{p,D}^2 = 4\pi n_D e^2/m^\ast$ is the square of the Drude plasma
frequency, $n_D$ is a carrier concentration associated with coherent transport,
$m^\ast$ is an effective mass, $\Gamma_D=1/\tau_D$ is the scattering rate, and
$\sigma_{MIR}$ is the mid-infrared component.  (When $\omega_{pD}$ and
$1/\tau_D$ are in cm$^{-1}$, $\sigma_1$ has the units $\Omega^{-1}{\rm
cm}^{-1}$.) The conductivity in the mid-infrared is often described by a series
of overdamped Lorentzian oscillators which yield a flat, incoherent response in
this region. To simplify the fitting and reduce the number of parameters, a
constant background ($\sigma_{MIR}\sim850$ $\Omega^{-1}{\rm cm}^{-1}$) has been
used. The fitted Drude parameter values are summarized in
Table~\ref{tab:drude}.
%
%

Between room temperature and 60~K, the normalized Drude carrier density (last
column of Table~\ref{tab:drude}) remains constant, while the scattering rate
scales with the temperature ($\hbar/\tau \simeq 2k_BT$), similar to what has
been seen in the normal state of other metallic cuprates \cite{gao93,rome92}.
Below 60~K, a new and surprising behavior occurs: the carrier density
decreases rapidly, with a reduction of nearly 75\%\ at 5~K.  We note that
this behavior is correlated with the onset of charge-stripe order as detected
by diffraction techniques \cite{fujita04,abba05}.  The scattering rate also
continues to decrease as $n_D$ drops.

It is interesting to compare with measurements of the in-plane resistivity,
$\rho_{ab}$, shown in the inset of Fig.~\ref{fig:sigma}(b), obtained on a
sister crystal.  On cooling, the decrease in $\rho_{ab}$ is interrupted by a
slight jump at the structural transition, at about 54~K
\cite{fujita04,ichikawa00}. The drop in $\rho_{ab}$ below about 40~K is
likely due to filamentary superconductivity \cite{moodenbaugh88,ichikawa00},
as magnetic susceptibility measurements show that the bulk superconducting
transition is at 2.4 K. The non-bulk response can be suppressed through the
application of a magnetic field (9~T); the resulting curve shows a nearly
constant value for the resistivity below 54~K until the onset of bulk
superconductivity at low temperature. The Drude expression for the dc
resistivity, in units of $\Omega\,{\rm cm}$, is $\rho_{dc} =
60/(\omega_{p,D}^2 \tau_D)$. The results obtained from the parameter values
in Table~\ref{tab:drude}, indicated by circles in the inset figure, are in
good agreement with the transport data.  Thus, the flattening out of
$\rho_{ab}$ and slight upturn at low temperature corresponds to a decrease in
carrier density, and not to an increase in scattering rate.

%
%
\begin{table}[t]
\caption{Drude parameters from two-component fits to the low-frequency
{\it ab}-plane conductivity of La$_{1.875}$Ba$_{0.125}$CuO$_4$.  The last
column relies on the assumption that $m^\ast$ is temperature independent.
}
\begin{ruledtabular}
\begin{tabular}[c]{cccc}%
T (K) & $\omega_{p,D}$ (cm$^{-1}$) & $1/\tau_D$ (cm$^{-1}$)
      &  $n_D(T)/n_D(60~{\rm K})$ \\%
\cline{1-4}
 295 & $7010\pm 120$ & $657\pm 35$ & $0.98\pm 0.12$ \\
 150 & $7070\pm  80$ & $223\pm  9$ & $0.99\pm 0.06$ \\
  60 & $7090\pm  50$ & $108\pm  2$ & $1.00\pm 0.06$ \\
  50 & $6560\pm  50$ &  $89\pm  2$ & $0.85\pm 0.05$ \\
  40 & $5740\pm  40$ &  $68\pm  2$ & $0.65\pm 0.04$ \\
  30 & $4930\pm  40$ &  $43\pm  2$ & $0.48\pm 0.04$ \\
   5 & $3690\pm  50$ &  $28\pm  2$ & $0.27\pm 0.04$ \\
\end{tabular}%
\end{ruledtabular}%
\label{tab:drude}%
\vspace*{-0.2cm}
\end{table}

\begin{figure}[t]%
%
%
\vspace*{-0.5cm}%
\centerline{\includegraphics[width=3.7in]{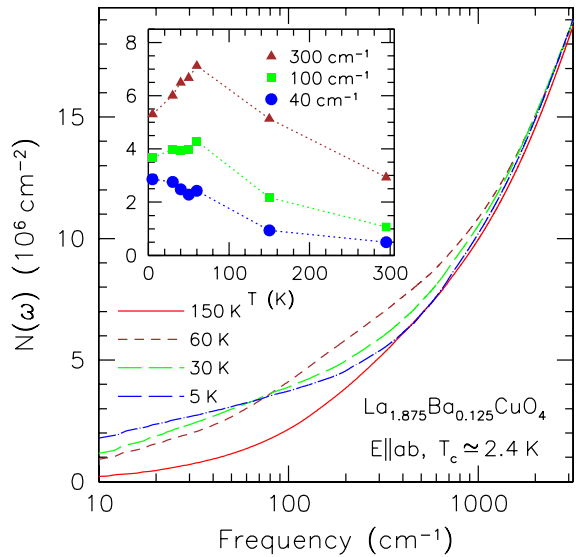}}%
\vspace*{-1.4cm}%
%
%
%
\caption{(Color online) The {\it ab}-plane spectral weight N$(\omega)$
of La$_{1.875}$Ba$_{0.125}$CuO$_4$ for several different temperatures
above and below 60~K.  To simplify the units, the conductivity has been
expressed in cm$^{-1}$. Inset: The temperature dependence of the
spectral weight for cut-off frequencies $\omega_c = 40$, 100 and
300~cm$^{-1}$.}
\vspace*{-0.3cm}%
\label{fig:weight}
\end{figure}

%
%
The optical conductivity lost from the Drude peak must shift to a different
frequency range.  Where does the missing spectral weight go?  To see this,
we consider the spectral weight, given by
\begin{equation}
  N(\omega_c, T) = \int_{0^+}^{\omega_c} \sigma_1(\omega, T)\, d\omega .
\end{equation}
as a function of the cut-off frequency, $\omega_c$.  If $\omega_c$ is large
enough to cover all relevant transitions, then $N(\omega_c)$ should be
proportional to the carrier density and independent of the scattering rate;
this is known as the $f$-sum rule.  Figure~\ref{fig:weight} shows $N(\omega)$
for several temperatures above and below 60~K; the inset shows the
temperature dependence of $N(\omega_c)$ for three different values of
$\omega_c$.  As the temperature decreases from 295~K down to 60~K,
there is an increase in the low-frequency spectral weight, consistent with
results for La$_{2-x}$Sr$_x$CuO$_4$ \cite{orto05}.  On cooling below 60~K,
the spectral weight below 100~cm$^{-1}$ continues to increase \cite{sumrule};
however, the net spectral weight up to 300 cm$^{-1}$ shows a substantial
decrease.  The missing spectral weight is finally recovered for $\omega_c
\sim 2000$~cm$^{-1}$, above which $N(\omega)$ is temperature independent.
Similar behavior of the spectral weight has been observed in an
electron-doped cuprate system \cite{zimmers05}.  The loss of spectral weight
below 60~K suggests a gapping of states near the Fermi level, while the
continued presence of a Drude component indicates that some states are not
gapped. From these results we infer the presence of an anisotropic charge
excitation gap in the stripe-ordered phase.

%
%

It has been predicted theoretically that CDW order should compete with
superconductivity within an ordered array of stripes \cite{kivelson98}.
If this were the case, then, as discussed by Zhou {\it et al.}\
\cite{zhou99}, one would expect the development of a CDW gap within the
stripes to occur in the ``antinodal'' regions of reciprocal space, where
the extrapolated Fermi surface has straight, well-nested portions.  Note
that the antinodal states exhibit a pseudogap in the normal state
\cite{timusk99,yosh03}, so that development of a true CDW gap would only
affect states at finite binding energy.

Alternatively, we find that the decrease in the low-frequency
conductivity at low temperature in the stripe-ordered state of
La$_{1.875}$Ba$_{0.125}$CuO$_4$ looks surprisingly similar to that found in the
superconducting state of La$_{1.85}$Sr$_{0.15}$CuO$_4$ \cite{tajima05}, both in
terms of energy scale and magnitude.  A residual Drude peak was also observed
in the latter superconductor.  This similarity suggests an intimate connection
between the charge gap of the stripe-ordered state and the pairing gap of the
superconductor.  Could the charge gap correspond to pairing
without phase coherence?

In either case, the residual Drude peak indicates that states in the
``nodal'' region are not strongly impacted by the charge gap, so that the
stripe-ordered state remains a nodal metal \cite{lee05}.  We note that
the coexistence of metallic behavior and an anisotropic gap has been
observed in two-dimensional CDW systems
\cite{forro98,mcconnell98}.  Such behavior is contrary to expectations of
a uniform gap for ordered stripes \cite{castellani00}.

Before concluding, we should consider how our results compare with others in
the literature.  The two-component behavior (temperature-dependent Drude peak
plus temperature-independent mid-infrared component) has been identified
previously in studies of several different cuprate families
\cite{rome92,gao93,lee05}. The loss of spectral weight in the stripe ordered
phase is a new observation. Measurements on
La$_{1.275}$Nd$_{0.6}$Sr$_{0.125}$CuO$_4$ \cite{dumm02} seem to be consistent,
but a larger low-temperature scattering rate in that sample masks any
depression of the optical conductivity.   In contrast, studies by Lucarelli
{\it et al.}~\cite{lucarelli03} on La$_{2-x}$Sr$_x$CuO$_4$ and Ortolani {\it et
al.}~\cite{orto05b} on La$_{1.875}$Ba$_{0.125-y}$Sr$_y$CuO$_4$ have found
strong, sharp peaks at finite frequency ($20-100$~cm$^{-1}$) in low-temperature
measurements. These ``far-infrared peaks'', which have not been observed by
other groups, have been interpreted as collective excitations of charge stripes
\cite{lucarelli03,orto05b,benf03}. Those results have been controversial, and
have motivated some discussion of possible spurious effects
\cite{tajima03,lucarelli03b,tajima05}.  Here we simply note that our
measurements on a cleaved sample do not show any indications of far-infrared
peaks.  Furthermore, the temperature-dependence of the conductivity and the
loss of the low-frequency spectral weight that we observe is directly
correlated with onset of stripe order as detected by diffraction
\cite{fujita04,abba05}; the behavior of the Drude peak is quantitatively
consistent with the measured resistivity.

%
%
In summary, the {\it ab}-plane properties of a cleaved single crystal of
La$_{1.875}$Ba$_{0.125}$CuO$_4$ have been examined over a wide frequency
range at several different temperatures.  The rapid decrease of the
carrier concentration $n_D$ below the electronic transition at $\simeq
60$~K suggests the development of an anisotropic charge gap associated
with the stripe order.  Nodal excitations presumably remain ungapped; the
deviation from $\rho_{ab}\sim T$ at small $T$ is due to the decrease in
carrier density, not to a uniform increase in scattering rate.  Thus it
appears that stripes are compatible with the nodal-metal state.

%
%
\begin{acknowledgments}
We would like to acknowledge useful discussions with D.~N.~Basov and
T.~Timusk, and helpful comments from S.~A.~Kivelson. This work was supported
by the Office of Science, U.S. Dept. of Energy, under contract number
DE-AC02-98CH10886.
\end{acknowledgments}

%
%
%
\bibliography{lbco}

\begin{thebibliography}{35}
\bibitem[{sum()}]{sumrule}
\bibinfo{note}{The lower bound starts at zero frequency for the sum rule, and
  as a result the low-frequency extrapolation is included; this introduces an
  element of uncertainty as to the nature and position of the crossover. For
  instance, if the lower bound is chosen to be 10~cm$^{-1}$, the crossover
  becomes difficult to observe.}
\end{thebibliography}

\end{document}